\documentstyle[draft]{mn}
\begin{document}
\title[Optical imaging of nova remnants]{Deep optical imaging 
of nova remnants II. A southern-sky sample}
\author[C.D.\ Gill \& T.J.\ O'Brien ]
{C.D.\ Gill \& T.J.\ O'Brien 
\\
Astrophysics Research Institute, 
Liverpool John Moores University, Byrom Street,
Liverpool L3 3AF\\
e-mail: {\tt cdg} \& {\tt tob} {\tt @astro.livjm.ac.uk}}

\maketitle
\begin{abstract}
We present an optical imaging study of 20 southern-sky nova remnants
which has resulted in the discovery of four previously unknown nova
shells -- V842 Cen, RR Cha, DY Pup and HS Pup.  The study has also
revealed previously unobserved features in three other known shells --
those of BT Mon, CP Pup and RR Pic. The images of BT Mon, V842 Cen, RR
Cha, DY Pup and HS Pup have been processed using several deconvolution
algorithms (Richardson-Lucy, maximum entropy and clean) in addition to
straightforward point-source subtraction in an attempt to resolve the
shells from the central stars. The use of four different methods
enables us to make a qualitative judgement of the results. Notably,
the shell of RR Pic displays tails extending outwards from clumps in
the main ejecta similar to those previously detected in DQ Her.
\end{abstract}


\section { Introduction }
In paper I (Slavin, O'Brien \& Dunlop 1995) we presented
the results of a deep optical imaging survey of the nebular remnants of a
sample of northern-sky classical novae. This paper reports on the
extension of this survey into the southern-sky using observations made
with the Anglo-Australian Telescope in 1995. 

Classical nova eruptions result in the ejection of
$\sim10^{-4}\,M_{\odot}$ of material at velocities of up to several
thousand kilometres per second (Bode \& Evans 1989).  Every nova
should be surrounded by an expanding cloud of ejecta and therefore it
is perhaps surprising that the literature contains few images of the
nova shells themselves.  We therefore embarked on an imaging survey of
nova remnants taking advantage of the advances in detector technology
and image processing techniques that allow us to detect fainter and
smaller shells. Since this survey was undertaken we have also been awarded
time on the Hubble Space Telescope to image a selected subset of the novae
discussed here and in paper I. These data will be presented elsewhere 
along with the initial results from our kinematical study of the shells. 

In the next section we describe the observations and the image processing 
techniques employed to investigate any extended structure. In section 3 we 
go on to discuss the observations for each nova for which we detect
extended material. Basic data and 
background information for each of the novae are obtained from Duerbeck (1987)
and Bode \& Evans (1989). We conclude with 
a section on estimating distances via expansion parallax and a brief 
discussion of the implications of these results. 

\section { Observations and image processing}
The observations were carried out at the Anglo-Australian Observatory on
the 3.9-m Anglo-Australian Telescope on 1995 February 23--25. We used
TAURUS-2, the Fabry-Perot spectrometer, in direct imaging mode with a
Tektronix CCD detector at f/15 providing a pixel scale of 0.32~arcsec.
All images presented in this paper were taken with an
H$\alpha$/[N{\scriptsize II}]
filter of central wavelength 6555$\rm \AA$ and full width at half maximum
54$\rm\AA$. Table~\ref{novaetab} summarizes the observations.
Surface brightness estimates of the shells were not  attempted as the
observing conditions were non-photometric. Also nova shells are known to
emit in H$\alpha$ and [N II], which are both transmitted by the
filter used, making measurement of the separate line fluxes impossible.
In any case, our primary aim, to detect nova shells and to investigate
their morphologies, does not require photometry.

\begin{table*}
\begin{center}
\begin{tabular}{cccccc}
\bf {Nova}&	\bf {Integration}&\bf{$t_3$ time}	&\bf {Seeing}
&\bf {Extended?} &\bf Shell Size\\
	    &	\bf {time / s}	   &\bf {/ days}      	& \bf { FWHM / arcsec}
&\bf {(Yes/No)} &\bf / arcsec\\
V365 Car    	&900	&530	&1.7	&N	\\
MT Cen	    	&1800	&--	&1.0	&N	\\
V359 Cen	&2$\times$900	&--	&1.6	&N	\\
V842 Cen	&4$\times$400+200	&48    	&0.7	&Y	&$<1.5$\\     
RR Cha		&1800   &60	&0.8      	&Y	&3$\times$2\\
AR Cir		&1800	&415	&0.6	&N	\\
AP Cru		&2$\times$900	&--	&2.0	&N	\\	
BT Mon		&2$\times$900	&36 / $190^1$	&1.8	&Y	&11$\times$9\\
GI Mon		&2$\times$900	&23	&1.0	&N	\\
GQ Mus		&2$\times$900	&45	&1.8	&N	\\
IL Nor		&1800	&108	&0.6	&N	\\
V841 Oph	&2$\times$900	&130	&0.7	&N	\\
V972 Oph	&1800	&176	&2.0	&N	\\
RR Pic		&1800	&150	&1.1	&Y	&30$\times$21\\
CP Pup		&1800	&8	&1.2	&Y	&19.5\\
DY Pup		&1800	&160	&0.8	&Y	&7$\times$5\\
HS Pup		&1800	&65	&0.8	&Y	&$<2.5$\\
HZ Pup		&2$\times$900	&70	&1.7	&N	\\
XX Tau		&2$\times$900	&42	&1.6	&N	\\
CQ Vel		&1800	&50	&1.1	&N	\\
\\
\end{tabular}
\end{center}
\caption{The novae imaged at H$\alpha$/[N{\scriptsize II}] are listed
together with the integration times (the components of co-added images
are listed if appropriate), their speed class in terms of $t_3$ (the time
in days taken to decline by three magnitudes from peak brightness), the
seeing measured from each image in arcsec, whether the nova appeared to
be extended beyond a normal point-spread function and if so, its
approximate diameter. Note 1 -- the $t_3$ time for BT Mon is disputed in
the literature (see text). }
\label{novaetab}
\end{table*}

\begin{figure*}
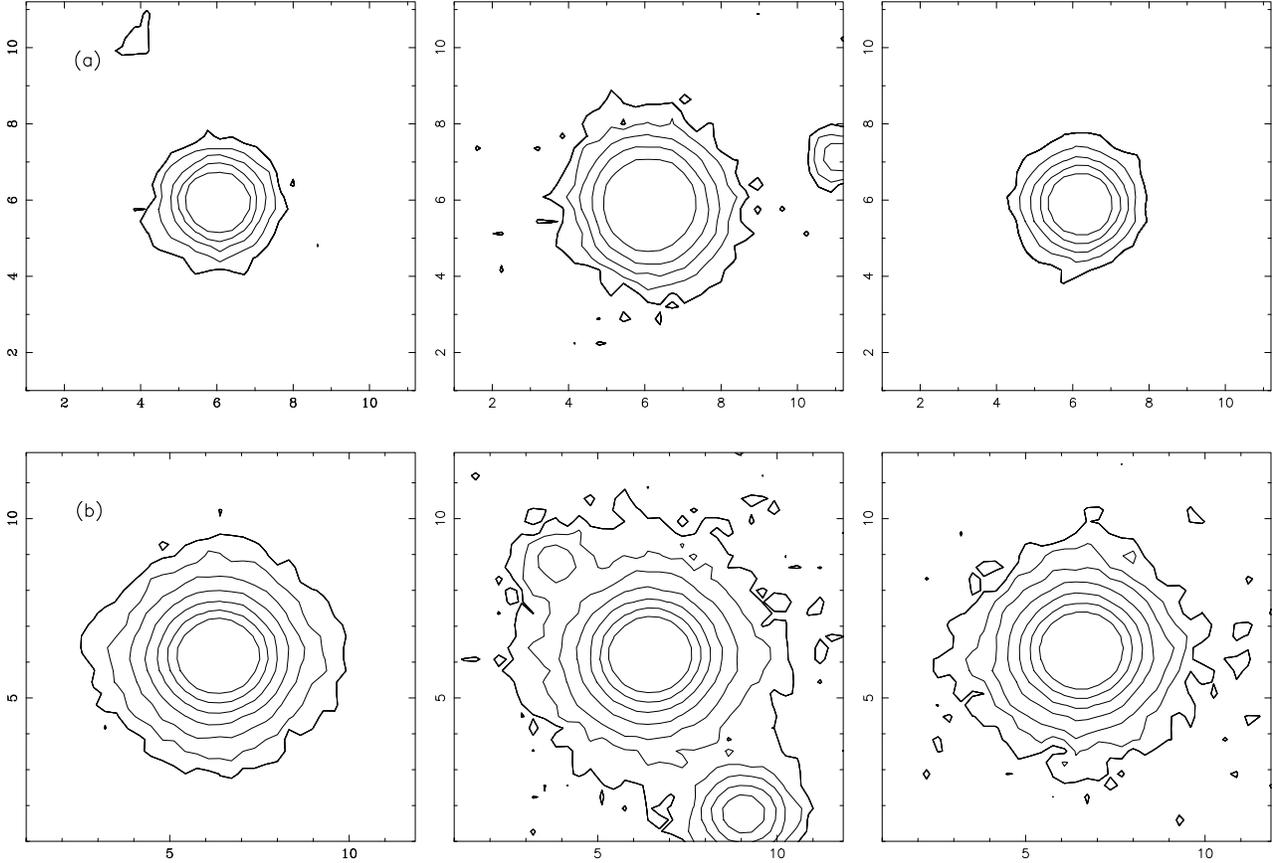

\vspace {11.5truecm}
\includegraphics{hspup_st.eps}
\includegraphics{gimon_st.eps}
\caption[]{ Top panel (a): 
Contour map of HS Pup (centre) and two scaled down brighter
stars from the same frame (on either side) displayed at contours of
0.128, 0.064, 0.032, 0.016 and 0.008 of the peak flux. Lower panel (b):
Contour map of
GI Mon (centre) and two scaled down brighter stars from the same frame
(on either side) displayed at contours of 0.128, 0.064, 0.032, 0.016,
0.008, 0.004 and 0.002 of the peak flux. The final contour in both sets
of images is two sigma above the sky background. 
Axes are marked in arcsec.}
\label{contours}
\end {figure*}

\begin{figure*}
\vspace {20.8truecm}
\includegraphics{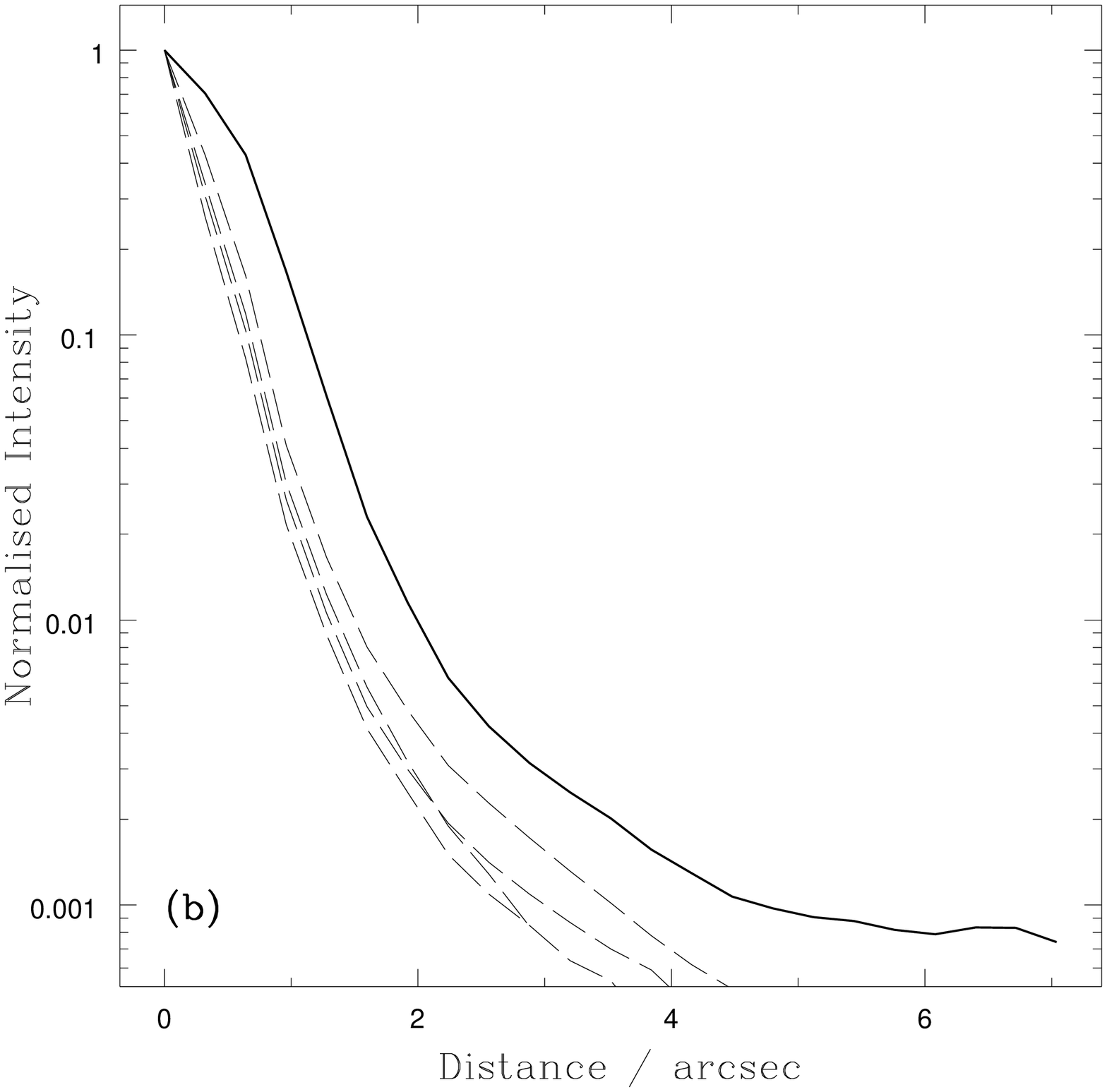}
\includegraphics{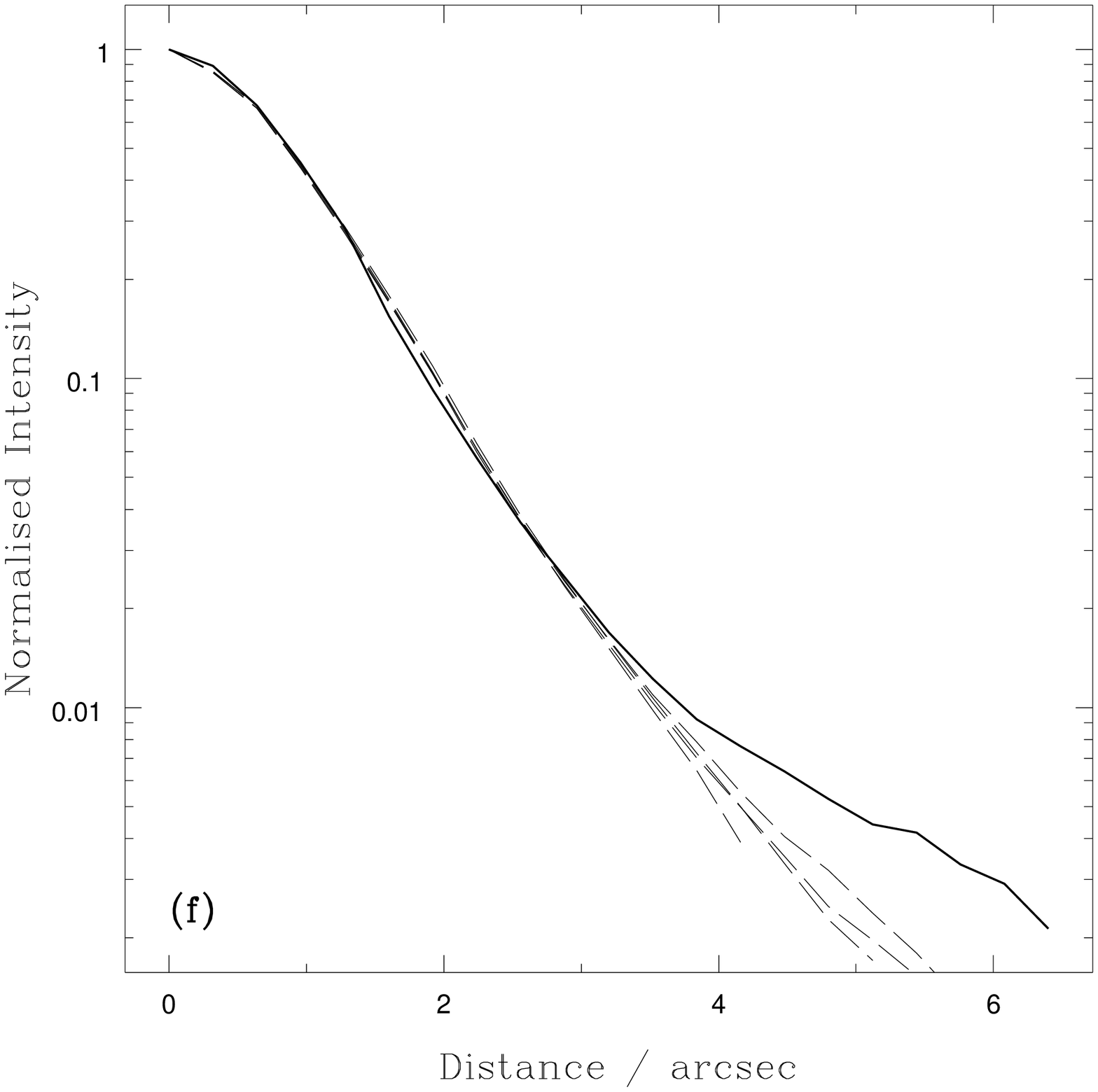}
\includegraphics{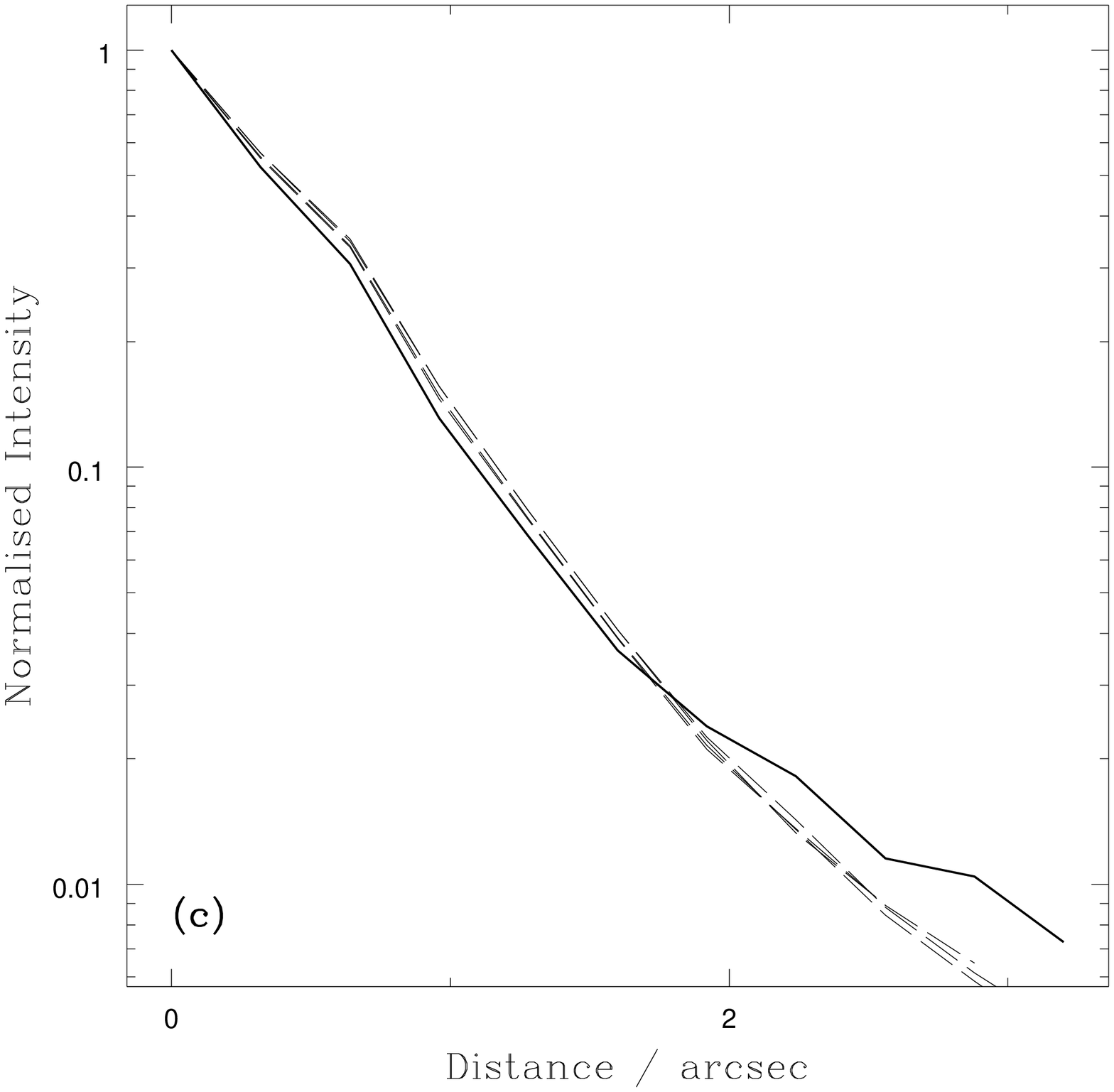}
\includegraphics{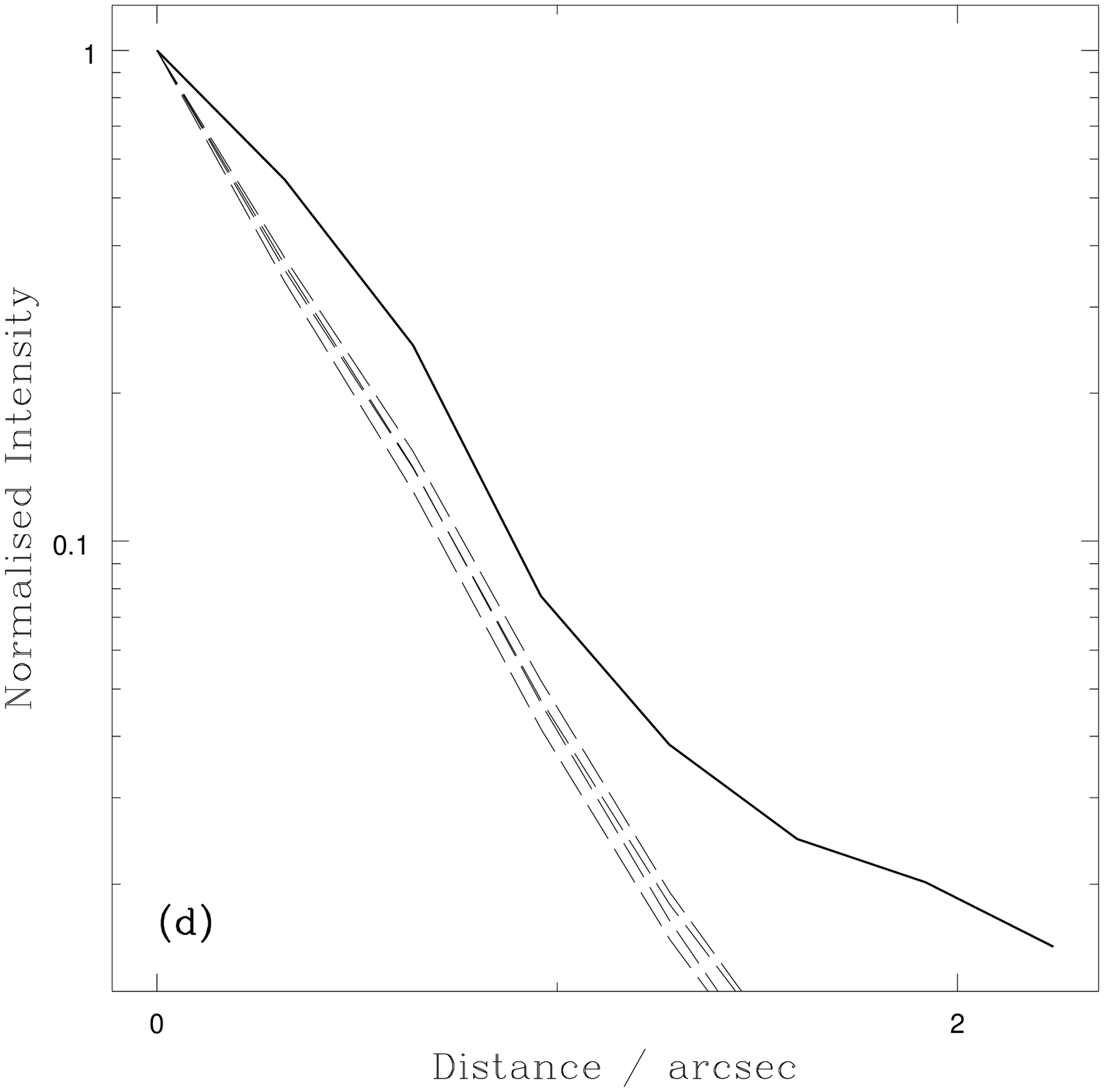}
\includegraphics{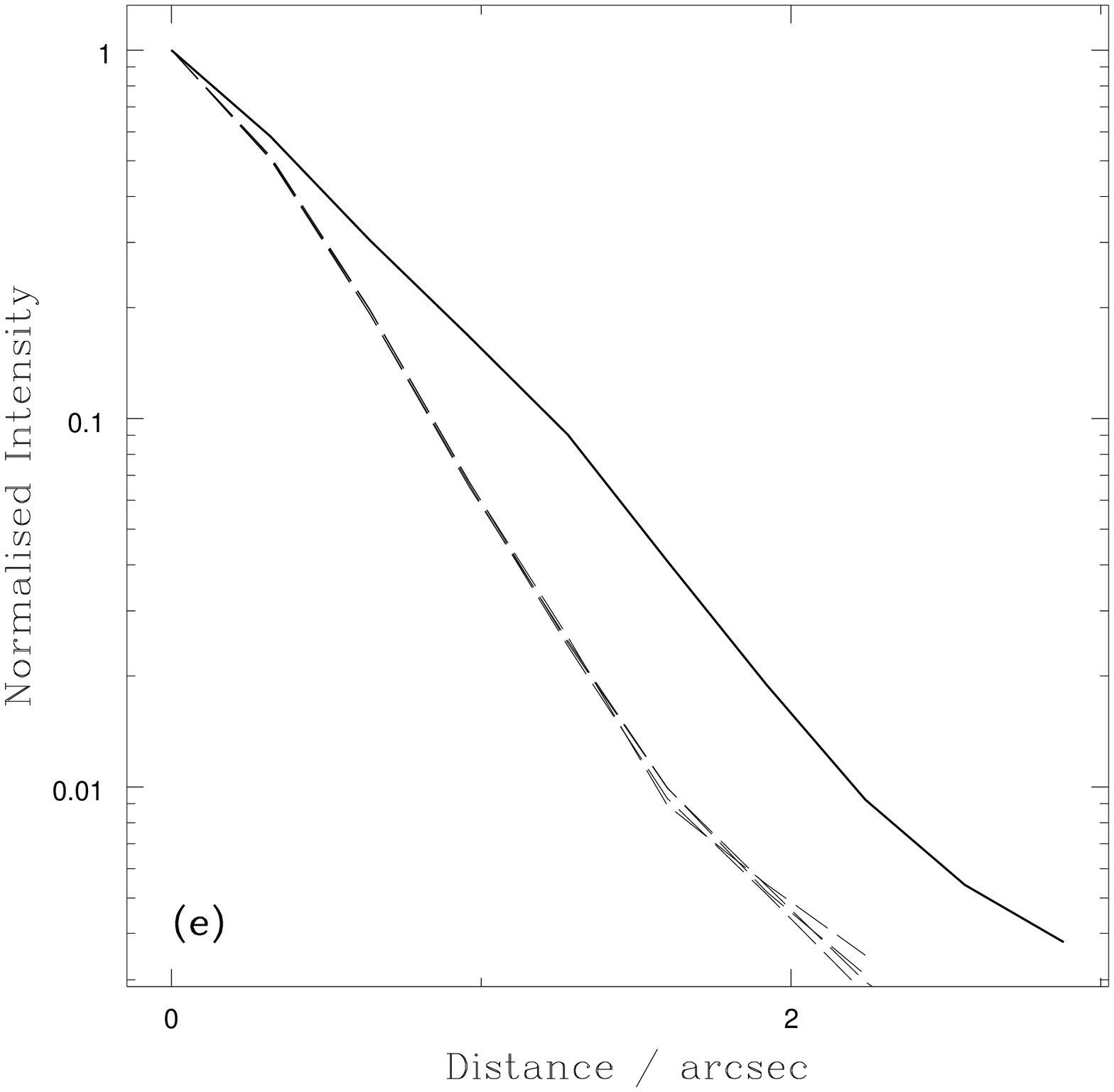}
\includegraphics{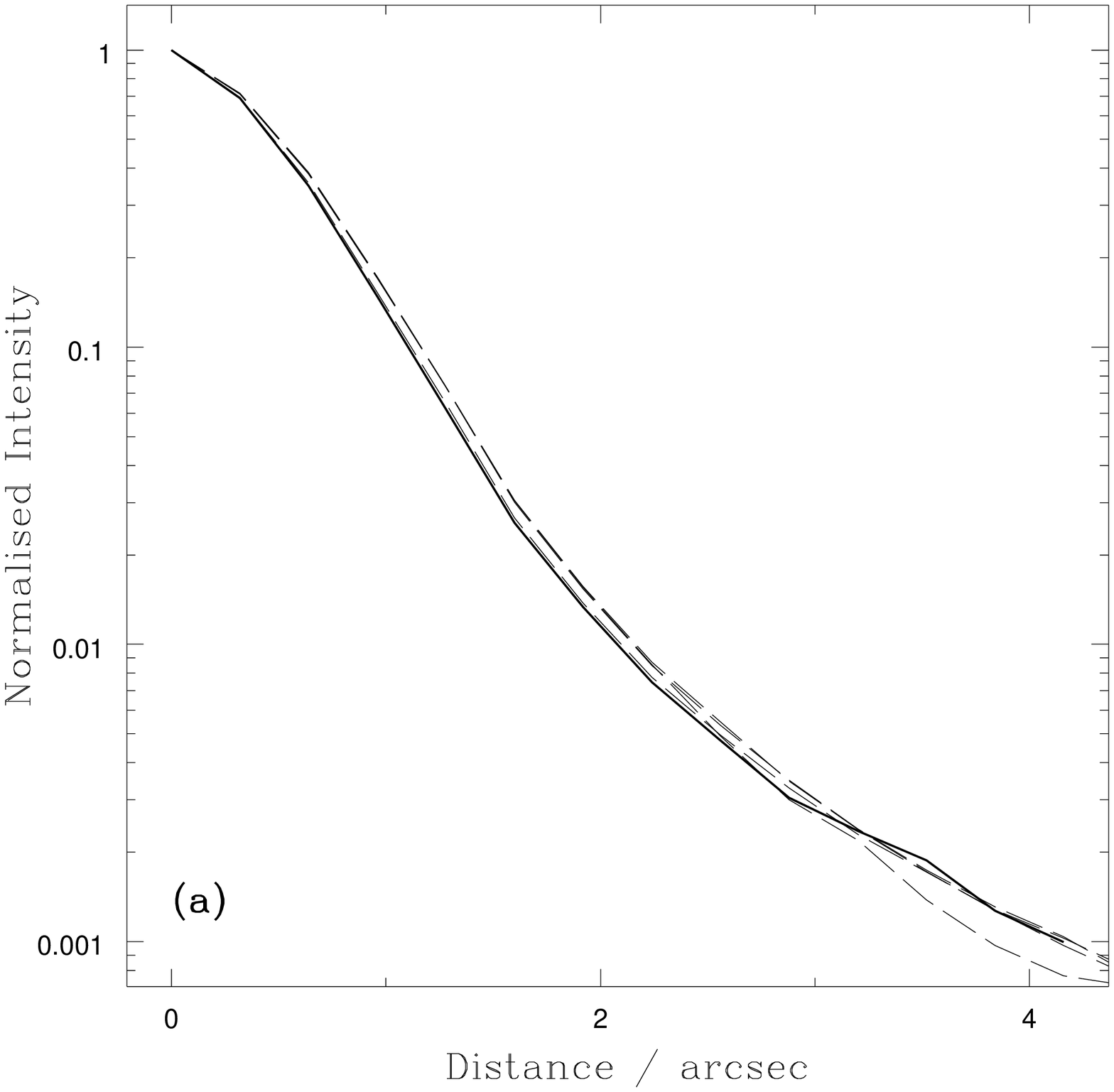}
\vspace {1.0truecm}
\caption[]{ Azimuthally-averaged brightness profiles for several images 
from our survey. In each case solid
lines designate the nova and dashed lines denote 4 field stars taken 
from the same
frame. All profiles are shown down
to a cut off one sigma above the sky.  
a) GI Mon -- no obvious extended material (shown for comparison).
b) V842 Cen -- clearly extended over the whole radial range. 
c) RR Cha -- the nova follows a stellar profile out to about 2 arcsec 
where some extended material becomes apparent. 
d) DY Pup -- apparently slightly extended over the whole range
but with a large increase at about 2 arcsec. 
e) HS Pup -- clearly extended at all distances shown. 
f) BT Mon -- no obvious extended material until a radius of about 4 arcsec.}
\label{profiles}
\end {figure*}

We investigated each image for any initial evidence of extended material using
two methods. We compared the nova to two stars taken from the same frame
using contour maps of images of each object scaled to the same peak
brightness. Figure~\ref{contours} shows the results for HS Pup, an object for
which we claim to have detected extended emission, and GI Mon, for which
there was no evidence of extension. We also present, in Figure~\ref{profiles},
azimuthally averaged brightness profiles for each of the novae described in 
section 4 for which we believe we have detected extended emission. 
In each case the nova light profile is compared
with several stars taken from the same frame. The only difficulty with these
methods is when the star and the nova are of significantly different
peak brightness, in which case one or the other will fade into the sky
whilst the other is still detectable. In order to avoid confusion due to
this effect we have presented our results for each method only up to a
radius where the star or nova brightness is more than 1 sigma above the
mean sky level.

If the methods described above led us to believe that extended
emission might be present then several techniques were
used to try to determine the morphology of this material in more
detail.  

The most basic of these was straight-forward star subtraction. We selected a 
star from the same frame as each nova (a brighter star was used to ensure 
adequate signal to noise in the wings of the point-spread function), this was
shifted to the same centroid as the nova, scaled to the same peak intensity
and subtracted from the nova image.

We also attempted to deconvolve the nova images using maximum entropy
({\sc mem}) and Richardson-Lucy ({\sc lucy}) routines available in the
UK Starlink Software Collection.  The final processing technique used
was our own {\sc clean} routine (for further details see Slavin,
O'Brien \& Dunlop 1994).
The use of a variety of processing techniques enables us to judge, at
least qualitatively, whether structure in the extended material is
real or simply an artefact of a particular algorithm. As can be seen
in the next section, the results from these techniques proved to be in
broad agreement.

\section {Images of the nova shells}

\subsection { V842 Centauri}

V842 Cen (Nova Cen 1986) was discovered on 1986 November 22.
Maximum light was missed but the $t_3$ time
(the time taken to decline three magnitudes from peak brightness) was
estimated to be 48 days by Sekiguchi et al.\ (1989) and it is therefore
classified as a moderately fast nova.

\begin{figure*}
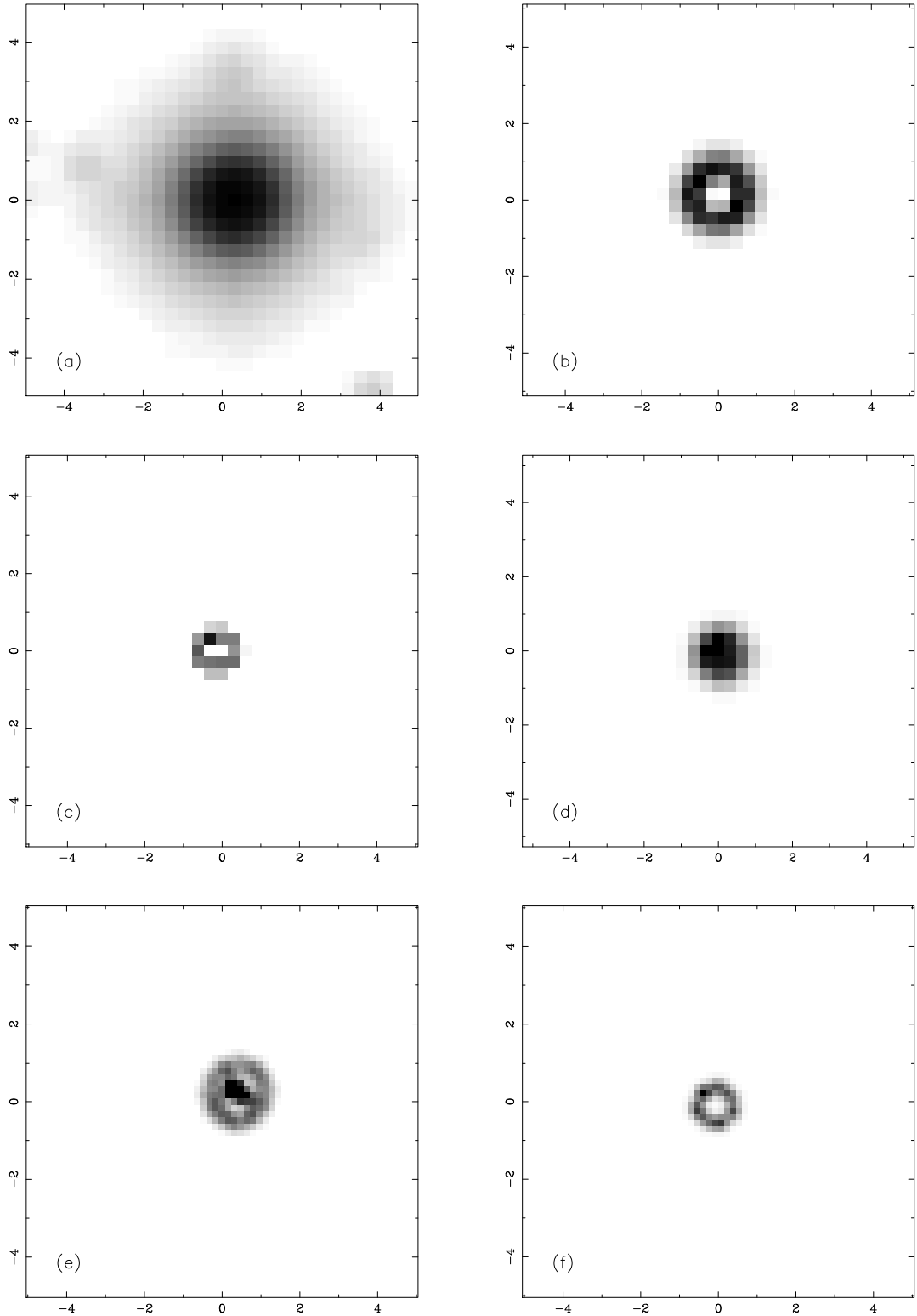

\vspace {20.8truecm}
\includegraphics{cen86_im.eps}
\includegraphics{cen86_cl.eps}
\includegraphics{cen86_ly.eps}
\includegraphics{cen86_me.eps}
\includegraphics{cen86_ss.eps}
\includegraphics{cen86_m2.eps}
\vspace {1.0truecm}
\caption[]{ H$\alpha$/[N{\scriptsize II}] images of the remnant of V842
Cen. In all five plots north is up, east is to the left and the axes are
marked in arcseconds. The images are: 
a) the original frame shown as a logarithmically grey scaled image, 
b) the star-subtracted image of the remnant, 
c) {\sc lucy} deconvolved image of the remnant, 
d) {\sc mem} deconvolved image of the remnant, 
e) {\sc clean}ed image of the remnant, and 
f) super-pixellated {\sc mem} deconvolved image of the remnant. }
\label{v842images}
\end {figure*}

To prevent saturation of the very bright central star, V842 Cen was
observed for 4 separate integrations of 400s and 1 integration of
200s. These were then co-added to form the final image shown in
Figure~\ref{v842images}(a).  By comparison with a star from the same
frame of slightly higher peak brightness it is clear that V842 Cen is
extended beyond a normal PSF, e.g.\ Fig.\ \ref{profiles}(b).

The results from the four image processing techniques are displayed in
Figure~\ref{v842images}(b-e). The star-subtracted image
(Fig.~\ref{v842images}b) shows a bright ring with radius of approximately
0.8~arcsec. The peak brightness of this ring is about 38\% of the peak
brightness of V842 Cen's central star and when a field star and the nova
are both scaled to the same peak brightness we find that the nova has a
total luminosity of twice that of the star. This is clear evidence for
bright, somewhat extended, emission-line material.

The {\sc lucy} deconvolution is shown in Fig.~\ref{v842images}(c) whilst
the result of the {\sc mem} deconvolution is displayed in
Fig.~\ref{v842images}(d). The {\sc lucy} deconvolution produces a bright
small ring comparable with the pixel scale whereas the {\sc mem}
deconvolution fails to resolve the ejecta from the central source,
producing a wide smooth single PSF. The {\sc clean} technique,
Fig.~\ref{v842images}(e), does deconvolve the nova into a central point
source and a ring. This latter technique employs 2$\times$
super-pixellation (linearly interpolating the image onto a grid with half
the original pixel scale) prior to deconvolution and the results are
smoothed with a 0.64 arcsec FWHM Gaussian. The {\sc clean} procedure is
bound to result in a central point source as this simply represents the
position of peak brightness in the original image. We believe that the
{\sc mem} method failed to resolve the shell because the shell is so
small in comparison to the seeing and the pixel scale. However if this
method is used on the super-pixellated data then it results in a ring but
no central star (Fig.~\ref{v842images}f) comparable to the result from
the {\sc lucy} deconvolution. The non-detection of a central source is
presumably because the shell is so bright and a ring of emission is the
simplest model consistent with the data. Clearly these results have to be
treated with some caution but as the shell is so bright, it is a good
candidate for observations with the Hubble Space Telescope or from the
ground in years to come as it increases in size and becomes easier to
resolve from the central star.

\subsection { RR Chamaeleontis }

Maximum light was also missed for RR Cha. It occurred between April 8 and
July 13 in 1953. However, the $t_3$ time has been estimated to be 60 days. 

\begin{figure*}
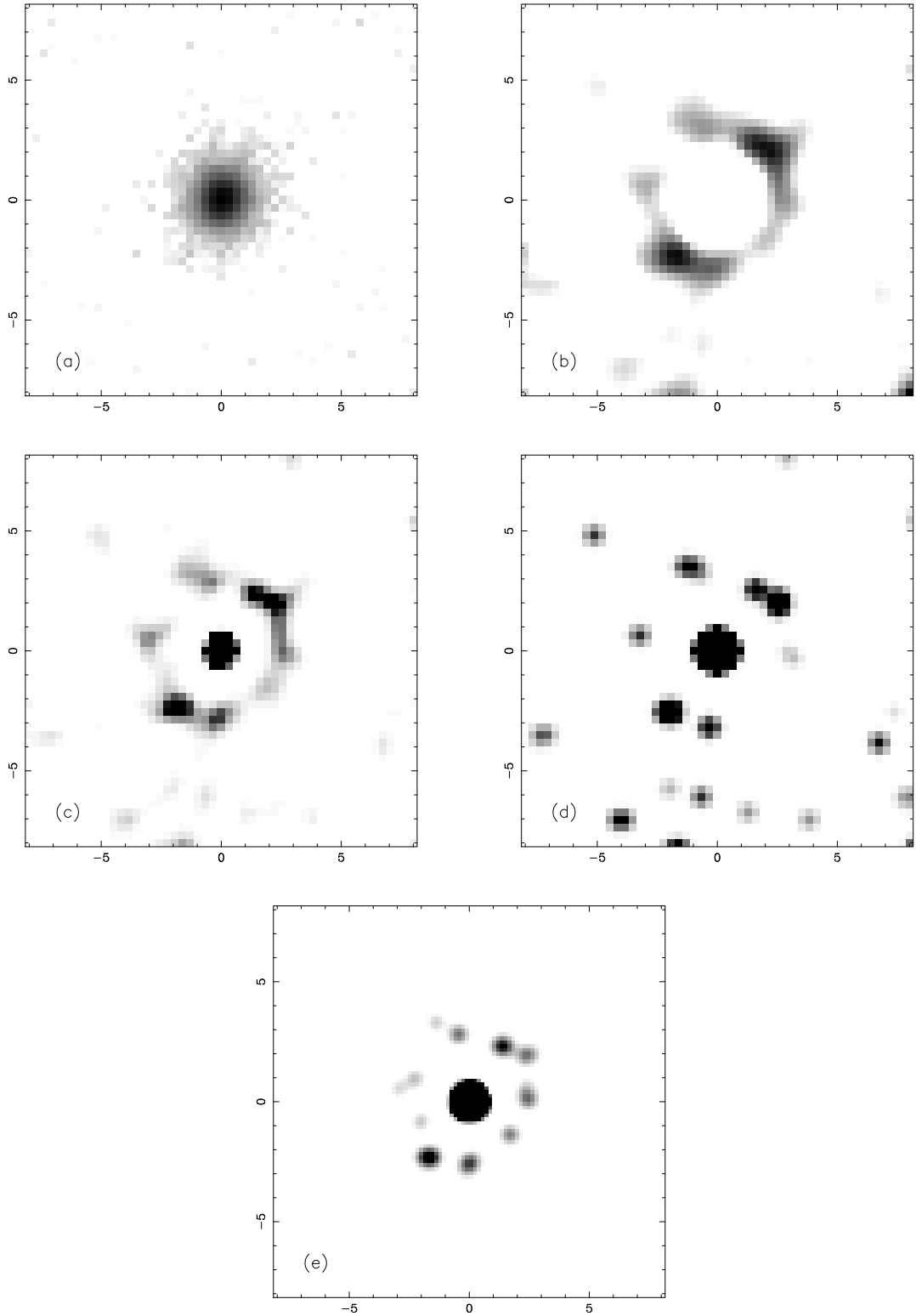

\vspace {20.8truecm}
\includegraphics{rrcha_im.eps}
\includegraphics{rrcha_cl.eps}
\includegraphics{rrcha_ly.eps}
\includegraphics{rrcha_me.eps}
\includegraphics{rrcha_ss.eps}
\vspace {1.0truecm}
\caption[]{ H$\alpha$/[N{\scriptsize II}] images of the remnant of RR
Cha. In all five plots north is up, east is to the left and the axes are
marked in arcseconds. The images are: a) the original frame shown as a
logarithmically grey scaled image, b) star subtracted image of the
remnant (smoothed with a 1.28 arcsec FWHM Gaussian), c) {\sc lucy}
deconvolved image of the remnant, d) {\sc mem} deconvolved image of the
remnant, and e) {\sc clean}ed image of the RR Cha remnant (smoothed with
a 0.64 arcsec FWHM Gaussian). }
\label{rrchaimages}
\end {figure*}

RR Cha was imaged for one integration of 1800s,
Fig~\ref{rrchaimages}(a).  This was possible due to the relative
faintness of the central star.  The star subtracted frame is shown in
Fig.~\ref{rrchaimages}(b). The results have been smoothed using a
1.28~arcsec Gaussian to allow faint residual emission to be seen
clearly above the noise. The extended material takes the form of an
elliptical ring with a major axis of 3~arcsec and minor axis
2~arcsec. The brightest parts of the ring are on the major axis.

The result of the {\sc lucy} deconvolution of RR Cha is shown in
Fig.~\ref{rrchaimages}(c). This looks very similar to the
star-subtracted image. The {\sc mem} deconvolution is shown in
Fig.~\ref{rrchaimages}(d). In this case the extended material takes
the form of blobs rather than a complete ring but these are located at
the same positions as the brightest parts of the ring seen in (b) and
(c).  A similar result is found from the {\sc clean} routine,
Fig.~\ref{rrchaimages}(e).

\subsection { BT Monocerotis }

BT Mon reached maximum in September 1939. The $t_3$ time is disputed
in the literature due to the exact date of maximum light being missed.
Payne-Gaposchkin (1957) quotes a $t_3$ of 36 days calculated from
spectral data whereas Schaefer \& Patterson (1983) quote a $t_3$ of 190 days
from the light curve after maximum. This relies on the nova having a
plateau of 8.5 magnitudes at maximum whereas Payne-Gaposchkin argues
that the spectra imply that BT Mon could have been as bright as mag.\
5.0.  To complicate matters Duerbeck (1987) describes BT Mon as a
`probably fast nova' quoting from Duerbeck (1981) a $t_3$ of 42 days.

BT Mon has been found to be an eclipsing system with a period of
$8^h\,1^m$ and therefore we are looking at the system approximately
edge-on.  The shell has been detected around this nova in spectroscopic
observations by Marsh et al.\ (1983). They determine a distance of 1800
pc to BT Mon assuming the shell, with a diameter of 7 arcsec, was
expanding at 1500 km~s$^{-1}$. The slit position used was running
east--west over the central star. No direct images of the nebula have
been reported that show any clear morphological structure.

\begin{figure*}
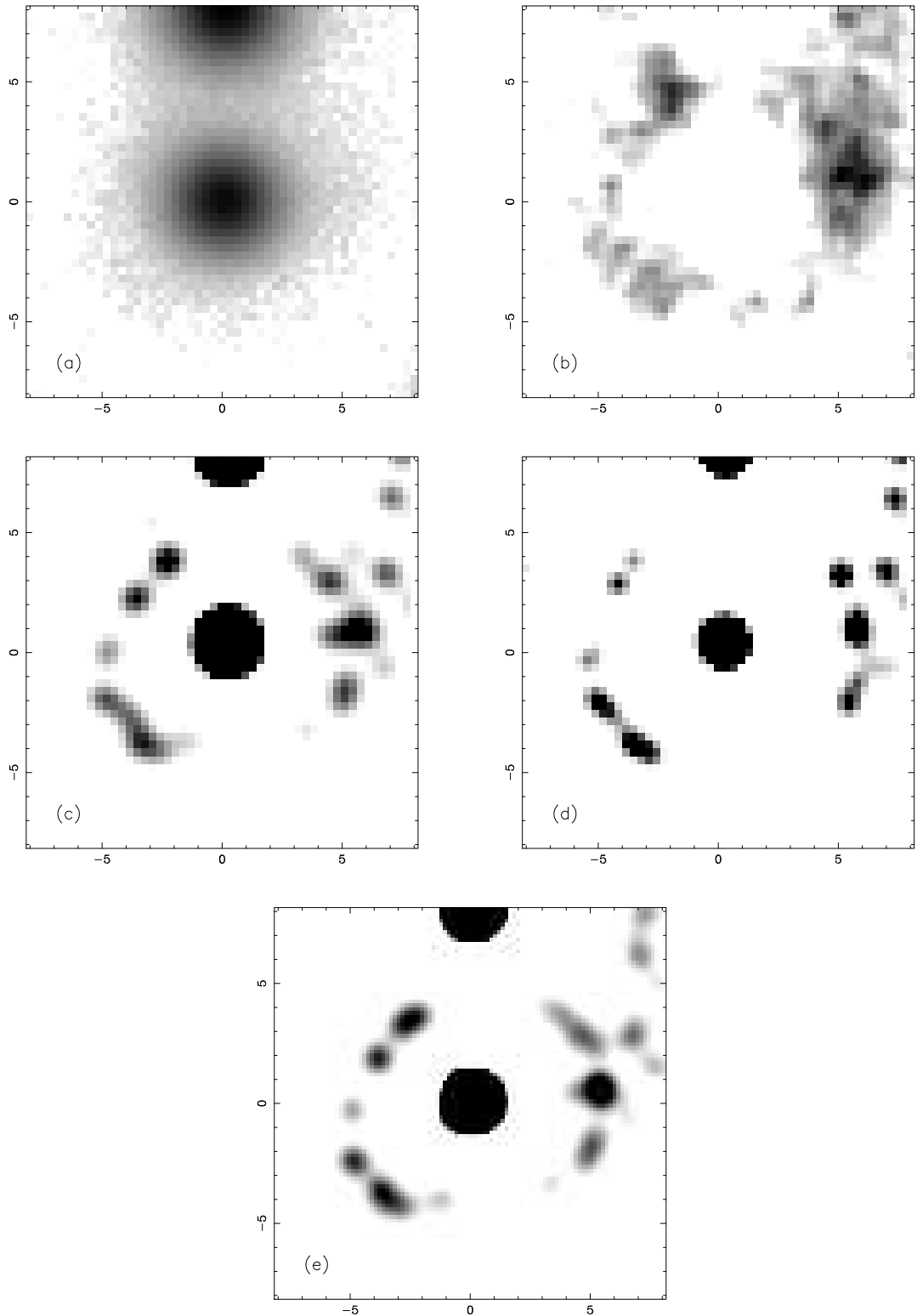

\vspace {20.8truecm}
\includegraphics{btmon_im.eps}
\includegraphics{btmon_cl.eps}
\includegraphics{btmon_ly.eps}
\includegraphics{btmon_me.eps}
\includegraphics{btmon_ss.eps}
\vspace {1.0truecm}
\caption[]{ H$\alpha$/[N{\scriptsize II}] images of the remnant of BT
Mon. In all five plots north is up, east is to the left and the axes are
marked in arcseconds. The images are: a) the original frame shown as a
logarithmically grey scaled image, b) star subtracted image of the
remnant (smoothed using a 0.64 arcsec Gaussian), c) {\sc lucy}
deconvolved image, d) {\sc mem} deconvolved image, and e) {\sc clean}ed
image (smoothed with a 0.64 arcsec FWHM Gaussian).}
\label{btmonimages}
\end {figure*}

Fig.~\ref{btmonimages}(a) shows the image of BT Mon resulting from
co-adding two 900s exposures. If one takes into account that the field
star to the north is about 30\% brighter than BT Mon itself then it is
possible to see that the wings of the BT Mon point spread function do
extend out further than those of this nearby star, indicating the
presence of some nebular material.

The results of star subtraction are displayed in
Fig.~\ref{btmonimages}(b). Both stars have been subtracted and the result
has then been smoothed using a 0.64
arcsec FWHM Gaussian (2 pixels). 
The {\sc lucy} result is shown in Fig.~\ref{btmonimages}(c), the {\sc mem}
result in (d) and the {\sc clean}ed image in (e). 
All processes show an incomplete clumpy, slightly elliptical ring with 
approximate dimensions of 11$\times$9 arcsec and a major axis extending 
in the north-west--south-east direction. 
The brightest material running east--west discovered by Marsh et al.\
(1983) in 1981 was found to have a diameter of 7 arcsec. As BT Mon
underwent eruption in 1939 then if we assume this material is moving at a
constant velocity then it should have a diameter of 9.3 arcsec in 1995,
when these observations were taken. The values measured from our
deconvolved images are about 10 arcsec which is in broad agreement 
with the predicted value.

\subsection { RR Pictoris }

\begin{figure*}
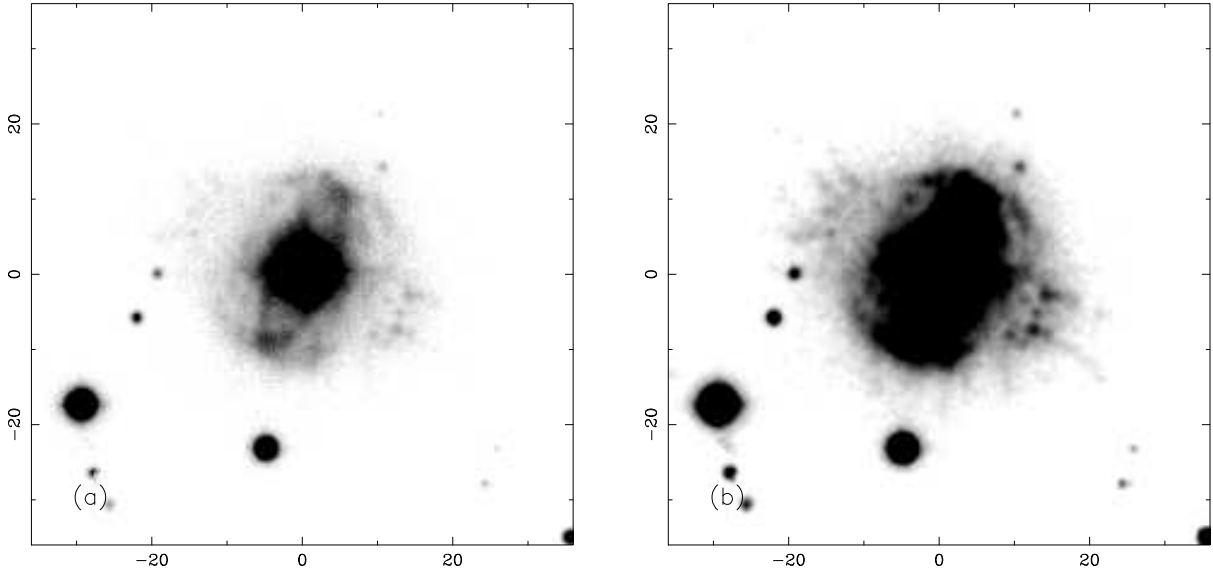

\vspace {7.0truecm}
\includegraphics{rrpic_hi.eps}
\includegraphics{rrpic_lo.eps}
\vspace {1.0truecm}
\caption[]{ H$\alpha$/[N{\scriptsize II}] images of the remnant of RR
Pic. North is up, east is to the left and the axes are marked in
arcseconds. The same image is displayed at different levels in order to
emphasise different features. In a) the bright equatorial ring is
apparent whilst in b) the fainter emission and tails orthogonal to the
major axis of the ring are more clearly seen. }
\label{rrpicimages}
\end {figure*}

RR Pic is a well-studied old nova. It reached maximum in 1925 with an
apparent visual magnitude of 1.0. It had a $t_3$ of 150 days which
characterises it as a slow nova. The period of the central system has
been measured at $3^h\,28^m\,50^s$

The extended shell has been known for a long time. The first material to
be detected was by van den Bos and Finsen in the early 1930's (see
Williams \& Gallagher (1979) and references therein) who observed bright knots
on opposite sides of the central system which they observed to be moving away
from the centre over 3 years with position angles (PAs) of 70$^\circ$ and
230$^\circ$ and a separation of 2 arcsec in 1931. Apparently the
next observations of the shell were made by Williams \& Gallagher (1979) who
again detected these knots. By this time they had a total separation of
23 arcsec implying a constant rate of expansion since 1931.
The latest published image of RR Pic found in the literature was in Evans
et al.\ (1992) (hereafter E92). The image therein resembles our own, shown in
Fig.~\ref{rrpicimages}(b), apart from the fact that our image appears to 
have much better spatial resolution.

The deconvolution techniques could not be safely used on the image of RR
Pic as there were no unsaturated stars in the frame brighter than the
nova central system itself. If a fainter star were used to characterise
the PSF then its wings would disappear into the noise before those of the
central system of RR Pic leading to spurious detection of extended
material. However, the shell is so large and clearly detected in our
images that deconvolution is not necessary.

The image of RR Pic is shown at two levels in
Fig.~\ref{rrpicimages}. Fig.~\ref{rrpicimages}(a) has been displayed
with scaling chosen so that the brighter equatorial ring can be
seen. This feature has a major axis of 21 arcsec at a position angle
of approximately 150$^\circ$.  Fig.~\ref{rrpicimages}(b) has been
displayed so that the fainter material and tails
running out to the north-east and south-west can be seen. O'Brien \&
Slavin (1996) investigated the tails using the same image processing
technique employed by Slavin, O'Brien \& Dunlop (1995).
An image taken through a [N{\scriptsize II}]6594$\rm \AA$ filter shows only the
southwest half of the equatorial ring indicating that this side of the shell
is tilted away from the observer (whether the emission is intrinsically
H$\alpha$ or [N{\scriptsize II}]).
 
Assuming a constant rate of expansion for the knots shown in Williams
\& Gallagher
(1979) then we expect them to have a total separation in 1995, when these
images were taken, of 30 arcsec. The separation of the knots on our image
is indeed 30 arcsec.

\subsection { CP Puppis}

CP Pup reached maximum light on 1942 November 9 and faded very quickly
with a $t_3$ time of 8 days classifying it as a very fast nova.  The
shell was first resolved by Zwicky (see Bowen 1956) and the latest image
found in a search through the literature was in Williams (1982). He shows
an H$\alpha$+[N{\scriptsize II}]6584\AA~ image taken at the CTIO
4m telescope in 1980 with an exposure of 4500 seconds onto a baked 098-04
plate which only just detects the clumpy shell of diameter 14
arcsec.

\begin{figure*}
\vspace {7.5truecm}
\includegraphics{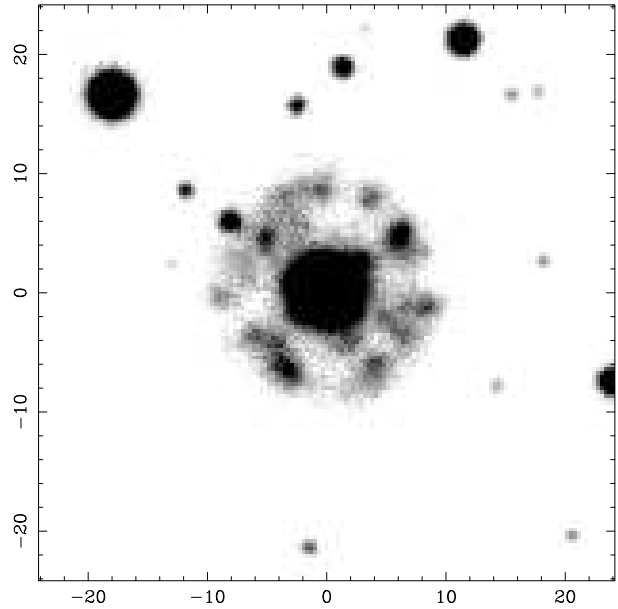}
\vspace {1.0truecm}
\caption[]{ H$\alpha$/[N{\scriptsize II}] image of the remnant of CP Pup.
North is up, east is to the left and the axes are marked in arcseconds. }
\label{cppupimage}
\end {figure*}

There has been no attempt to deconvolve the image of CP Pup, shown in
Fig.~\ref{cppupimage}, for the same reasons as RR Pic.
In Williams (1982) the peak to peak diameter of the shell in the NW to
SE and NE to SW directions is approximately 10 arcsec. If we assume
constant expansion since that image was taken in 1980 then the shell
size in our observation should be about 14 arcsec. The measured peak to
peak diameters in these directions are 15 and 14.5 arcsec
respectively which is in broad agreement with constant expansion.

\subsection { DY Puppis }

DY Pup reached maximum in November 1902 and declined slowly with a
$t_3$ of 160 days.  Little other work has been done on this nova with
no spectroscopic information available.

\begin{figure*}
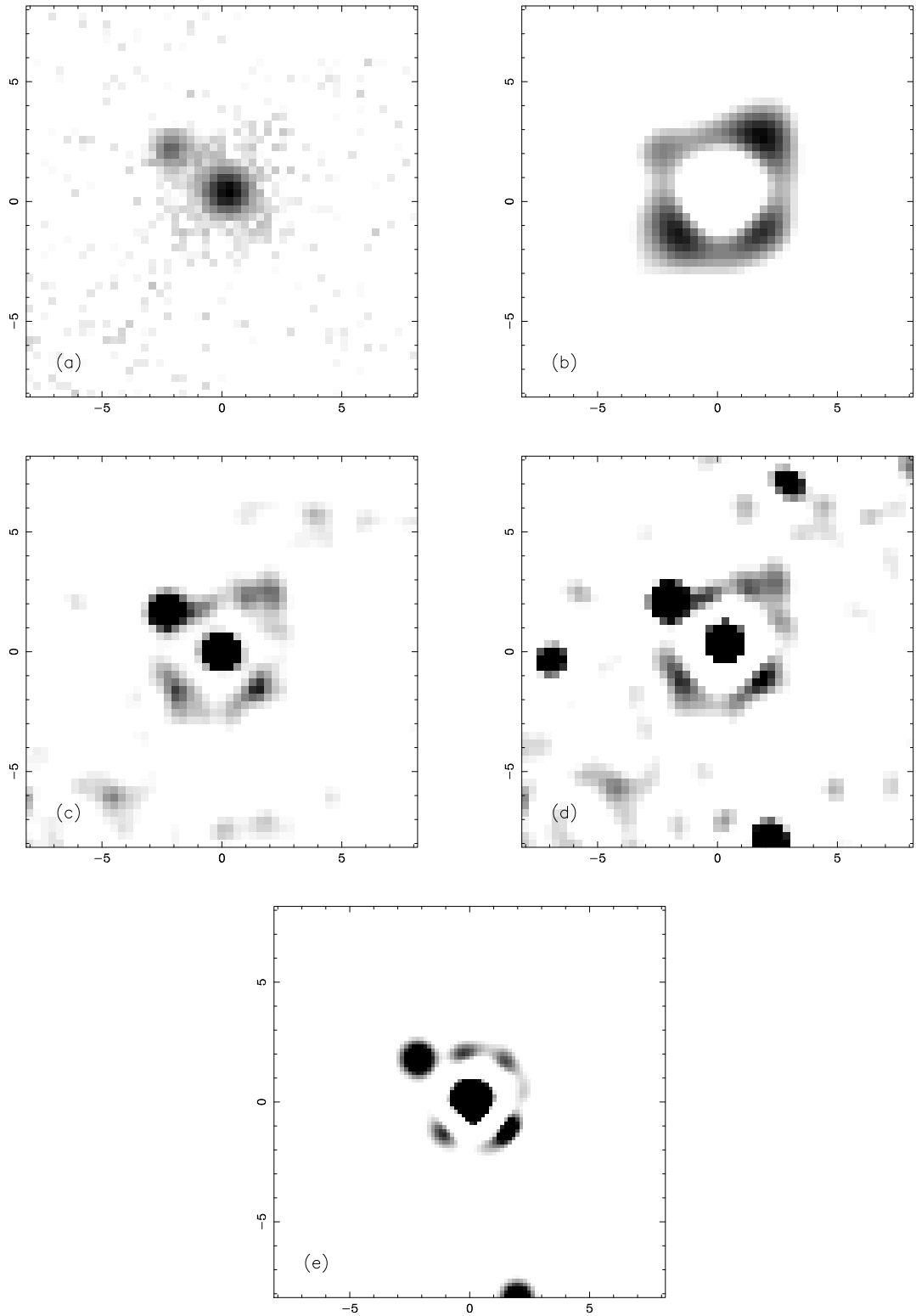

\vspace {20.8truecm}
\includegraphics{dypup_im.eps}
\includegraphics{dypup_cl.eps}
\includegraphics{dypup_ly.eps}
\includegraphics{dypup_me.eps}
\includegraphics{dypup_ss.eps}
\vspace {1.0truecm}
\caption[]{ H$\alpha$/[N{\scriptsize II}] images of the remnant of DY
Pup. In all five plots north is up, east is to the left and the axes are
marked in arcseconds. The images are: a) the original frame shown as a
logarithmically grey scaled image, b) star subtracted image of the
remnant (smoothed with a 2 arcsec FWHM Gaussian), c) {\sc lucy}
deconvolved image, d) {\sc mem} deconvolved image, and e) {\sc clean}ed
image (smoothed with a 0.64 arcsec FWHM Gaussian). }
\label{dypupimages}
\end {figure*}

The raw image of DY Pup and its surroundings resulting from a single
1800s integration is shown in Fig.~\ref{dypupimages}(a). It is
possible to make out some very faint extended material although the
object just to the north--east of DY Pup appears to be a field star.

The star subtracted image is shown in Fig.~\ref{dypupimages}(b). In
this case, both the central star and the nearby field star have been
subtracted and the residual smoothed using a 2 arcsec FWHM
Gaussian. The result is a ring of material with enhancements to the
north-east, north-west, south-east and south-west. It appears to be slightly 
elongated in the north-west--south-east direction.
The {\sc lucy} deconvolved image is shown in Fig.~\ref{dypupimages}(c).
The two stars are picked out clearly leaving an extended ring around DY
Pup bearing a close resemblance to the results of star subtraction.
The {\sc mem} result for DY Pup, Fig.~\ref{dypupimages}(d), reveals the same 
morphology as does the {\sc clean} result, shown in Fig.~\ref{dypupimages}(e).

\subsection { HS Puppis }

HS Pup underwent outburst in late December 1963. It reached maximum light
on the 23rd at an apparent photographic magnitude of 8.0. It decreased in
brightness at a moderately fast rate with a $t_3$ of 65 days. There is little
data on the nova since then other than its light curve.

\begin{figure*}
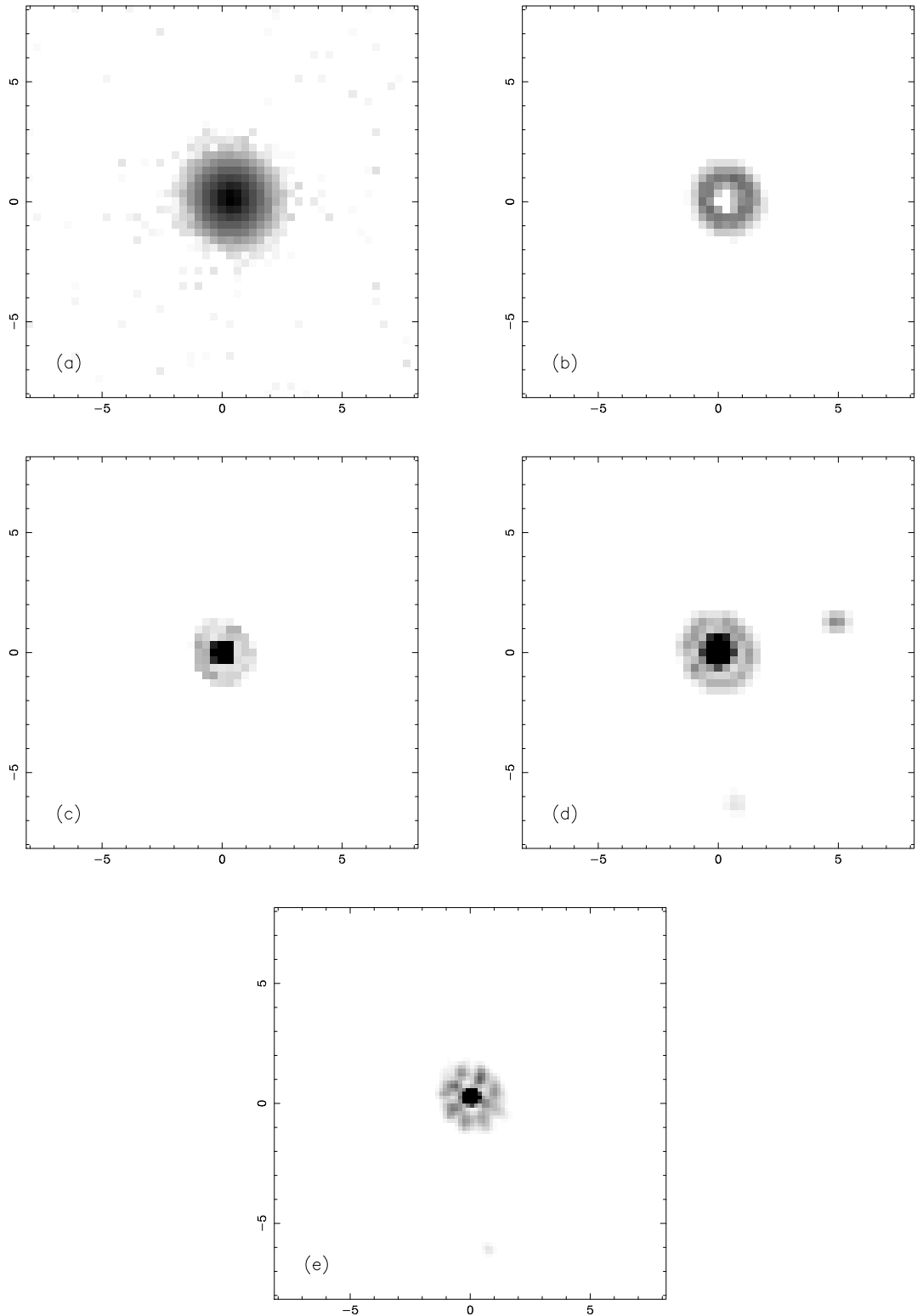

\vspace {20.8truecm}
\includegraphics{hspup_im.eps}
\includegraphics{hspup_cl.eps}
\includegraphics{hspup_ly.eps}
\includegraphics{hspup_me.eps}
\includegraphics{hspup_ss.eps}
\vspace {1.0truecm}
\caption[]{ H$\alpha$/[N{\scriptsize II}] images of the remnant of HS
Pup. In all five plots north is up, east is to the left and the axes are
marked in arcseconds. The images are: a) the original frame shown as a
logarithmically grey scaled image, b) star subtracted image of the
remnant, c) {\sc lucy} deconvolved image of the remnant, d) {\sc mem}
deconvolved image of the remnant, and e) {\sc clean}ed image (smoothed
with a 0.64 arcsec FWHM Gaussian).}
\label{hspupimages}
\end {figure*}

The original image resulting from a single 1800s exposure is shown in
Fig.~\ref{hspupimages}(a). There is no obvious extended material but,
like V842 Cen, when compared with a star scaled to the same peak
value, its brightness profile extends out a lot further from the
centre of the star (Fig.~\ref{profiles}e). 

The star subtracted image is displayed in
Fig.~\ref{hspupimages}(b). The {\sc lucy} deconvolution is shown in
Fig.~\ref{hspupimages}(c). It has resolved the object into a central
star and a surrounding halo of radius approximately 1 arcsec.  The
{\sc mem} deconvolution, Fig.~\ref{hspupimages}(d), also resolves the
object into a central star surrounded by a halo of extended material
which more closely resembles a detached ring in this case of
approximately 1.2~arcsec radius.  Fig.~\ref{hspupimages}(e) shows the
result of the {\sc clean} algorithm.  The results are similar to (b)
and (d) although the ring does appear clumpier.  This is almost
certainly a result of the discrete nature of the {\sc clean} process
and should not be regarded as a definite property of this remnant.

\section{Distance Estimates}

Given that we have obtained estimates of the angular extent of several nova 
shells it is possible, at least in principle, to make distance estimates based
on the method of expansion parallax using measurements of the ejection
velocity made during outburst. 
 If velocities ($v$) are given in km s$^{-1}$ , angular sizes ($\theta$) in
arcsec and times since ejection ($t$) in years then the distance ($d$) 
in parsecs is given by $d\,=\,0.211\,vt/\theta$.
In all possible cases the ejection velocity is estimated from the 
separation of symmetric spectral features. This means that systemic
velocities are not important although in any case these will be small compared 
with the ejection velocities.

The distance to V842 Cen has been calculated by Sekiguchi et
al. (1989) using the equivalent widths of interstellar Na I D lines in
their spectra. They derived a distance of $d=0.9 \pm 0.07$ kpc. The
measured diameter of the shell in our deconvolved images is about 1.6
arcsec. The time since eruption when these images were taken was 9
years. Sekiguchi et al. (1989) obtained spectra of the Balmer lines
for H$\alpha$ to H$\eta$ and saw a two peaked structure in all of
the lines. The average split between the peaks was 1070 km
s$^{-1}$. Assuming that these peaks were produced by the expanding 
shell we derive an expansion parallax distance of 
$d=1.3 \pm 0.5$ kpc. The large errors
are due to the large uncertainty in the diameter of the shell and the
expansion velocity.

Marsh et al. (1983) obtained a distance estimate to BT Mon of $d=1.8 \pm 0.3$
kpc from their long slit spectra. The expansion velocity they measure for
the shell is 1800 km s$^{-1}$. Taking the diameter of the shell to be 10
arcsec and the time since ejection to be 56 years we also derive a distance to
BT Mon of $d=1.8 \pm 0.3$ kpc. These estimates are limited by the poor velocity
resolution of Marsh and higher resolution spectra combined with our imaging 
data would enable a better
estimate to be made.

Payne-Gaposchkin (1957) adopts a value of 1600 km~s$^{-1}$ for the
velocity of the ejecta of CP Pup. If we take this value along with
a shell diameter in 1995 of 19.5 arcsec then the derived distance is
$d=1.8 \pm 0.4$ kpc. The uncertainty arises from assuming 
the velocity of the shell is known to an accuracy of 400 km $s^{-1}$. 

McLaughlin (1960) derives a distance to RR Pic of 480 pc from the angular
expansion of the knots of the ejecta (taken from Williams and Gallagher
1979). Payne-Gaposchkin (1957) presents results for the time evolution of a 
variety of different
lines. There are a consistent set of velocities
through the decline at $\pm 400$ km $s^{-1}$. Assuming the equatorial
ring is in fact circular then we derive an inclination angle of 
70$\pm 5^\circ$. Correcting the observed line of sight ejection
velocity for this angle we get an intrinsic velocity of 850~km~$s^{-1}$. 
The observed major axis of the ring was 21 arcsec so we then derive a 
distance to
RR Pic of $600 \pm 60$ pc. This is slightly larger than that of
McLaughlin as it also takes into account the inclination angle 
of the aspherical shell.

We could find no published information on expansion velocities for 
RR Cha, DY Pup and HS Pup and so no distance determinations were possible.

The distances derived in this section are uncertain for two obvious reasons -- 
the uncertainties in our
estimates of shell diameters and the measurements of ejection velocities 
obtained
from the literature. However, the discussion of RR Pic also raises the 
important point that distance estimates based on expansion parallax in the
 manner we describe here generally 
assume spherically symmetric ejection i.e.\ the 
ejection velocity is obtained from doppler shifts along the line of sight
whilst shell diameters are obtained from expansion in the plane of the sky. 
It is clear from these images and those presented in Slavin et al.\ (1995) that
many nova shells are far from spherical. Thus more accurate measures of 
distance will require some knowledge of the geometry of the shell. This can 
only be derived by combining spatially-resolved spectroscopy and imaging. In 
papers to follow we present the results of such spectroscopy/imaging 
for several of the more extended shells in our northern-sky sample.  

\section{Discussion}

Of the 17 novae without previously detected shells which were imaged 4
new shells were detected. The new material discovered was not resolved
from the central systems in any of these cases although scaled contour
maps (see Fig.~\ref{contours}) and brightness profiles (see
Fig.~\ref{profiles}) clearly proved that they were extended. To
investigate the morphology of these shells star subtraction, 
Richardson-Lucy, {\sc mem} and {\sc clean} deconvolutions 
were used. Input PSFs to
the routines were stars from the same frames which had brighter peak
intensities than those of the central systems of the novae to allow
proper sampling of the wings of the PSFs. The results of these
different methods were in general agreement allowing some confidence
in the predicted shell structures. Notably, extended tails similar to
those previously detected in DQ Her (Slavin et al.\ 1995) have been
discovered in the shell of RR Pic (see also O'Brien \& Slavin 1996).

Further analysis of, for example, the bipolarity of these shells is
not advisable due to the limitations of ground-based imaging of such
small angular-scale structures. However Hubble Space Telescope
observations of several of these objects are scheduled and will
hopefully improve on these data although it should be noted that the
observations in this paper are relatively long exposures using a
4m-class telescope. High-resolution
spectroscopy of these objects should also be obtained which, 
when combined with these images will enable distance estimates via
expansion parallax as well as further insight into the detailed
structure of the more extended objects.

\section {Acknowledgements}
We thank Keith Taylor and Joss Bland-Hawthorn for support at the 
Anglo-Australian Observatory,
Brent Tully for the loan of filters and Andy Slavin for
discussions regarding the northern-sky survey. CDG acknowledges support
from a PPARC studentship.

\end{document}